\documentclass[aps,twocolumn,showpacs,prl,superscriptaddress,preprintnumbers,amsmath,amssymb]{revtex4-1}

\usepackage{graphicx}
\usepackage{dcolumn}
\usepackage{bm}
\usepackage{braket}

\usepackage[usenames]{color}
\usepackage{ulem}

\begin{document}

\preprint{APS/123-QED}

\title{Zero-field Skyrmions with a High Topological Number in Itinerant Magnets}

\author{Ryo Ozawa}
\affiliation{Department of Applied Physics, University of Tokyo, Tokyo 113-8656, Japan}
\author{Satoru Hayami}
\affiliation{Department of Physics, Hokkaido University, Sapporo 060-0810, Japan}
\author{Yukitoshi Motome}
\affiliation{Department of Applied Physics, University of Tokyo, Tokyo 113-8656, Japan}

\date{\today}
\begin{abstract}
Magnetic skyrmions are swirling spin textures with topologically protected noncoplanarity.
Recently, skyrmions with the topological number of unity have been extensively studied in both experiment and theory. 
We here show that a skyrmion crystal with an unusually high topological number of two is stabilized 
in itinerant magnets at zero magnetic field.
The results are obtained for a minimal Kondo lattice model on a triangular lattice by an unrestricted large-scale numerical simulation and variational calculations. 
We find that the topological number can be switched by a magnetic field as $2\leftrightarrow 1\leftrightarrow 0$. 
The skyrmion crystals are formed by the superpositions of three spin density waves induced by the Fermi surface effect, and hence, the size of skyrmions can be controlled by the band structure and electron filling. 
We also discuss the charge and spin textures of itinerant electrons in the skyrmion crystals which are directly obtained in our numerical simulations.
\end{abstract}

\maketitle
\begin{figure*}[t]
	 \includegraphics[width=0.92\textwidth]{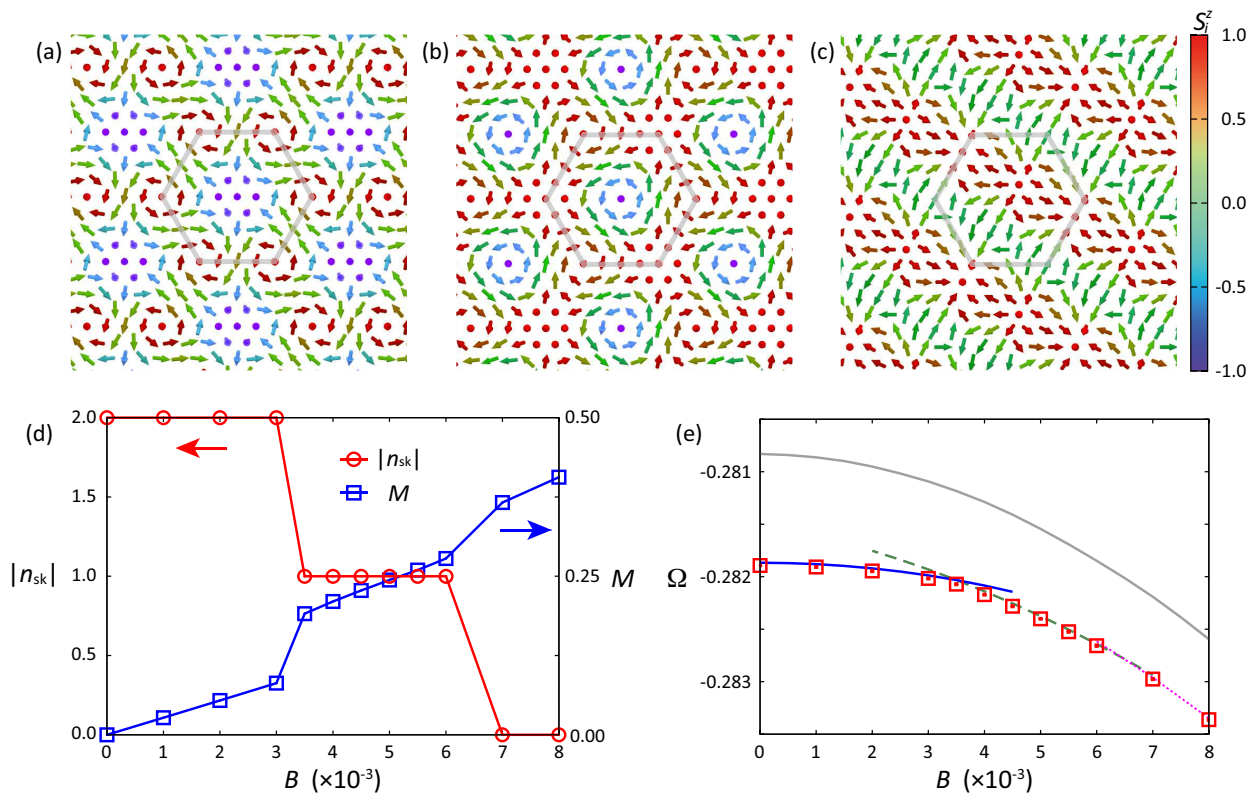}
	 \caption{
	 Configurations of localized spins in
	 (a) the $n_{\rm sk}=2$ SkX at $B=0$, 
	 (b) the $n_{\rm sk}=1$ SkX at $B=0.005$,  and
	 (c) the $n_{\rm sk}=0$ state at $B=0.008$.
	 (d) $B$ dependences of $|n_{\rm sk}|$ and $M$ obtained by the modified KPM-LD simulation. 
	 (e) Grand potential $\Omega$ obtained by the modified KPM-LD simulation (red squares),  
	 in comparison to those by the variational states for 
 	 the $n_{\rm sk}=2$ SkX in Eq.~(\ref{eq:Qsk=2}) (blue solid lines),
 	 the $n_{\rm sk}=1$ SkX in Eq.~(\ref{eq:Qsk=1}) (green dashed lines), 
 	 the $n_{\rm sk}=0$ state in Eq.~(\ref{eq:Qsk=0}) (magenta dotted lines), 
	 and 
	 the single-$Q$ conical state (gray solid lines) 
	 with optimal variational parameters at each $B$.
	 The modified KPM-LD results are obtained for the Kondo lattice model with $t_3 = -0.85$, $J=1.0$, and $\mu=-3.5$ on a triangular lattice with $N_s=96^2$.
	 The gray hexagons in (a)-(c) represent the magnetic unit cell.
	}\label{fig:skx} 
\end{figure*}

\begin{figure*}[t]
	 \includegraphics[width=0.92\textwidth]{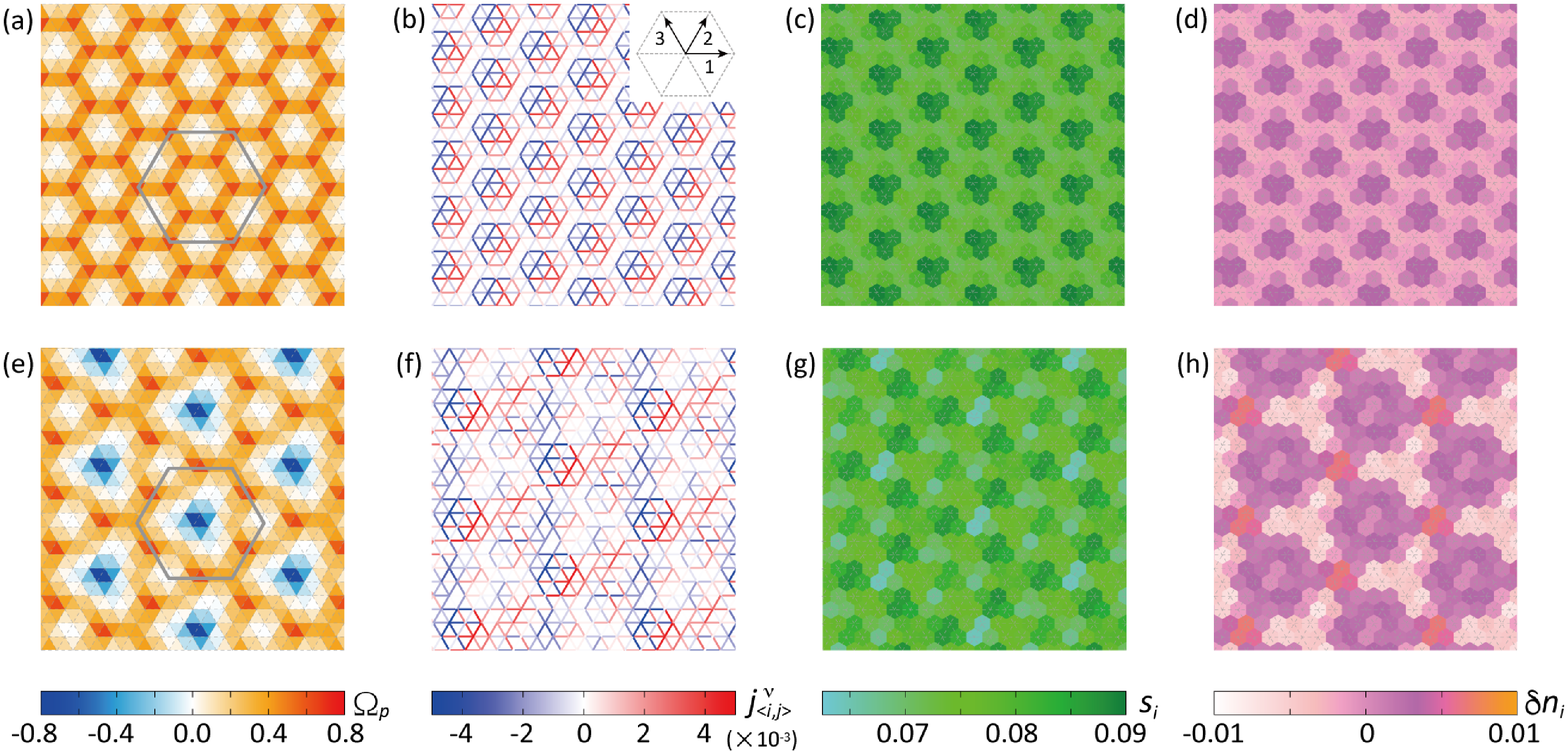}
	 \caption{ 
	 Real-space distributions of
	 (a) the solid angle of localized spins $\Omega_p$ on each plaquette, 
	 (b) the electric current density $j^\nu_{\langle i,j\rangle}$ on each bond,
	 (c) the spin length of itinerant electrons $s_i$ 
	 at each site,
	 and (d) the charge density $n_i$  
	 at each site measured from the average density 
	 for the $n_{\rm sk}=2$ SkX
	 obtained by the modified KPM-LD simulation at $B=0$. 
	 (e)-(h) Corresponding data for the 
	 $n_{\rm sk}=1$ SkX at $B=0.005$.
	 The other parameters are the same as in Fig.~\ref{fig:skx}. 
	 The gray hexagons in (a) and (e) represent the magnetic unit cell.
	 The inset of (b) shows the directions of the electric current densities, $\nu=1$, $2$, 
	 and $3$.
	 } 
	 \label{fig:electron}
\end{figure*}

The discovery of topological invariants in electronic states in solids has brought about 
an innovation in modern condensed matter physics.
For example, in the quantum Hall systems, the Chern number, which is given by the Berry connections of electron wave functions in momentum space, plays a key role in understanding of the quantized Hall conductivity~\cite{Berry, thouless1982quantized}. 
Another example found in magnetic systems is 
 the Pontryagin number, which is defined as the integral of the solid angles spanned by neighboring three spins, for characterizing the so-called magnetic skyrmions~\cite{muhlbauer2009skyrmion,yu2010real,nagaosa2013topological}.

The skyrmion is a swirling noncoplanar texture  
originally introduced as a hypothetical particle in the baryon theory~\cite{skyrme1962unified}. 
Since the experimental discovery 
of a skyrmion crystal (SkX) in the magnetically ordered state in MnSi~\cite{muhlbauer2009skyrmion},
the magnetic analog of the skyrmion has been extensively studied for not only the nontrivial topology but also the application to electronic devices~\cite{pfleiderer2010condensed, schulz2012emergent,iwasaki2013current,nagaosa2013topological,fert2013skyrmions,zhang2014magnetic,zhou2015dynamically}.
The noncoplanar spin texture in the  
SkX is described by a superposition of three helical spin density waves~\cite{muhlbauer2009skyrmion,yu2010real}, which is often stabilized by competition between the ferromagnetic exchange interaction and the Dzyaloshinskii-Moriya (DM) interaction originating 
 from the spin-orbit coupling.

Thus far, most of both experimental and theoretical studies have concerned with magnetic skyrmions 
with minimal Pontryagin number $n_{\rm sk}=1$ (called topological number hereafter). 
This is because higher $n_{\rm sk}$ costs an additional energy from larger relative angles between neighboring spins when the ferromagnetic interaction is dominant~\cite{belavin1975metastable}. 
However, skyrmions with any integer of $n_{\rm sk}$ are allowed in the topological sense, and 
rather desirable as they can bring about richer physics. 
For example, larger spin noncoplanarity can induce larger emergent electromagnetic fields for electrons, which may lead to larger responses in unconventional transport phenomena, such as the topological Hall effect~\cite{Berry, PhysRevLett.69.3232, PhysRevLett.83.3737, PhysRevLett.102.186602}.
Another advantage is that high-$n_{\rm sk}$ skyrmions may allow the multiple digital control of the topological numbers.
Nevertheless, stable high-$n_{\rm sk}$ skyrmions are very rare~\cite{yu2014biskyrmion,zhang2016high,leonov2015multiply}.
Moreover, the underlying stabilization mechanism remains elusive for  high-$n_{\rm sk}$ skyrmions, especially in thermal equilibrium.

In this Letter, we report our theoretical discovery of a thermodynamically-stable high-$n_{\rm sk}$ SkX.
The result is obtained for a minimal model for itinerant magnets, 
the Kondo lattice model with classical localized spins, on a triangular lattice, 
by using a sophisticated numerical method~\cite{PhysRevB.88.235101} 
 as well as variational calculations.
We find that the ground state shows a SkX with an unusual topological number $n_{\rm sk}=2$ at zero magnetic field, which has different spin textures from the previously-reported molecular-type biskyrmion~\cite{yu2014biskyrmion}. 
Furthermore, we find topological transitions with successive changes of 
$n_{\rm sk}$ from $2$ to $1$, and to $0$.
Interestingly, the size of skyrmions can be controlled by the band structure and electron filling, as the instability toward the SkXs is induced by the Fermi surface effect.
Our simulation also provides the properties of itinerant electrons in the SkXs, such as the electric current density, spin polarization, and charge density.

We consider the Kondo lattice model on a triangular lattice in a magnetic field. The Hamiltonian is given by
\begin{eqnarray}
\mathcal{H} =&-&\sum_{i,j, \sigma} t_{ij}^{\;}\hat{c}_{i\sigma}^\dagger\hat{c}_{j\sigma}^{\;}
- J\sum_{i}\hat{{\bf s}}_i\cdot{\bf S}_i - B\sum_i S^z_i,
\label{eq:model}
\end{eqnarray}
where the operator $\hat{c}_{i\sigma}^\dagger$ ($\hat{c}_{i\sigma}^{\;}$) creates (annihilates) an electron with spin $\sigma$ at site $i$.
The first term represents the kinetic energy of itinerant electrons
and $t_{ij}$ is the hopping integral between sites $i$ and $j$.
We consider the first- and third-neighbor hoppings,
$t_1$ and $t_3$, respectively, in the following calculations, while our results are expected to hold for generic cases as discussed later.
The second term describes the $s$-$d$ coupling between
itinerant electron spins $\hat{{\bf s}}_i = (1/2) \sum_{{\sigma\sigma'}}\hat{c}^\dagger_{i\sigma}{\bm{\sigma}}_{\sigma\sigma'}\hat{c}^{\;}_{i\sigma'}$
 and localized spins with coupling constant $J$,
where ${\bm  \sigma}$ is the vector of Pauli matrices.
For  simplicity, we treat the localized spins as classical vectors with unit length 
$|{\bf S}_i|=1$ (the sign of $J$ is irrelevant).
The third term denotes the Zeeman coupling to an external magnetic field, 
which is taken into account only for localized spins for simplicity. 
Hereafter, we take $t_1=1$ and $a=1$ (lattice constant)
as the energy and length units, respectively.

In the following, we examine the ground state of the model (\ref{eq:model})
in the weak $J$ region.
In the limit of $J\to 0$, the perturbation in terms of $J$ gives us an insight into anticipated spin ordering. 
The  lowest-order contribution is in the order of $J^2$, which is called the Ruderman-Kittel-Kasuya-Yosida (RKKY) interaction~\cite{PhysRev.96.99, Kasuya01071956, PhysRev.106.893}.  
The RKKY interaction is proportional to  the bare magnetic susceptibility 
$\chi^0_{\bf q} = - T\sum_{{\bf k},\omega_n} G^0_{{\bf k}, \omega_n}G^0_{{\bf k} + {\bf q}, \omega_n}$;
$T$ is the temperature and
$G^0_{{\bf k}, \omega_n}$ is the bare Green function of itinerant electrons with momentum ${\bf k}$
and Matsubara frequency $\omega_n$.
This apparently favors a helical spin ordering with 
the wave number at which $\chi_{\bf q}^0$ is maximized. 
In general, however, $\chi_{\bf q}^0$ has multiple maxima 
compatible with the lattice symmetry. 
For instance, in the triangular lattice case, there are at least three pairs of peaks 
reflecting the sixfold rotational symmetry. 
In such situations, the RKKY energy is degenerate for the helical orders with 
the symmetry-related wave numbers, and there arises a chance to have more complicated spin structures by superpositions of the helices. 
Indeed, a variety of such interesting spin textures have been found for the model (\ref{eq:model}) on several lattices~\cite{PhysRevLett.101.156402, JPSJ.79.083711, PhysRevLett.105.226403, PhysRevLett.108.096401, PhysRevLett.108.096403, PhysRevLett.109.166405, PhysRevB.90.060402}. 
Among them, the authors and coworkers found that the model (\ref{eq:model}) exhibits a noncoplanar vortex crystal composed of a superposition of two helices
as the generic ground state in the weak $J$ limit~\cite{JPSJ.85.103703}.

In this study, we explore further possibilities beyond the perturbative regime. 
For this purpose, we adopt an unrestricted numerical method based on the kernel polynomial method (KPM) and the Langevin dynamics (LD), 
which is called the modified KPM-LD method~\cite{PhysRevB.88.235101,preparation,ozawa2016shape}. 
This is highly efficient, whose computational cost is linear in the system size $N_s$, and enables us to treat large-scale systems of $N_s = 10^4$-$10^5$ by a massive parallel processing on GPUs. 
In the simulation, we perform the Chebyshev polynomial expansion up to $M = 2000$ with using $16^2$ correlated random vectors for the KPM and  set the time interval in  updating ${\bf S}_i$, $\Delta \tau=2$ in the LD for the systems with $N_s=96^2$.  

Figure~\ref{fig:skx}(a) shows the spin configuration obtained by the modified KPM-LD simulation at $B=0$ for the model with $t_3=-0.85$, 
 $J=1.0$, and the chemical potential $\mu=-3.5$.
For the parameters, $\chi^0_{\bf q}$ shows the peaks at commensurate wave numbers, ${\bf Q}_1=(\pi/3, 0)$ and ${\bf Q}_{2(3)}=R_{+(-)}{\bf Q}_1$, where $R_{+(-)}$ is a $(-)2\pi/3$-rotation operator~\cite{suppl}.
Remarkably, the optimal spin texture has a periodic array of a vortex with vorticity $v=-2$ centered at a downward spin surrounded by six vortices (merons) with $v=1$.

This is, to our knowledge, the first example of stable magnetic SkX with the topological number $n_{\rm sk}=2$ at zero magnetic field; 
here, $n_{\rm sk}$ is defined by the sum of solid angles spanned by neighboring localized spins over a magnetic unit cell~\cite{eriksson1990measure}.
Experimentally, a molecule of two bound skyrmions, called biskyrmion, was reported, but it appears in a nonzero magnetic field and the spin pattern is slightly different~\cite{yu2014biskyrmion}. 
It is, in general, hard to stabilize SkXs in the absence of magnetic field, while metastable ones were realized~\cite{okamura2016transition,kagawa2016quenching}. 
This is because the most of SkXs are with $n_{\rm sk}=1$, in which spins at the hull of each skyrmion are aligned in parallel, carrying a net magnetization. 
In stark contrast, our $n_{\rm sk}=2$ SkX has a spin configuration compatible with vanishing magnetization: the vorticity-two texture naturally results in a swirling but parallel spin configuration on the opposite side of the hull, giving zero net magnetization.
The $n_{\rm sk}=2$ SkX spontaneously breaks the chiral symmetry, 
while taking any value of the helicity in $[0:2\pi)$~\cite{nagaosa2013topological}.

We find that the $n_{\rm sk}=2$ SkX is well approximated by a superposition of three spin density waves (triple-$Q$) as
\begin{eqnarray}
{\bf S}^{(n_{\rm sk}=2)}_i\propto 
  (\cos\mathcal{Q}_{1i}, \cos\mathcal{Q}_{2i}, \cos\mathcal{Q}_{3i}),
  \label{eq:Qsk=2}
\end{eqnarray}
where $\mathcal{Q}_{\nu i} = {\bf Q}_\nu\cdot {\bf r}_i$  (${\bf r}_i$ is the position vector for site $i$).
This state is regarded as the generalization of triple-$Q$ states discussed in the previous studies at particular electron fillings~\cite{PhysRevLett.101.156402,JPSJ.79.083711}. 
In the previous cases, multiple-spin interactions are important, which arise from higher-order perturbation in terms of $J$ 
than the RKKY interactions~\cite{PhysRevLett.108.096401,PhysRevB.90.060402}. 
In the present case of the $n_{\rm sk}=2$ SkX,
we find that similar high-order perturbative terms 
play a role, as discussed later.

In an applied magnetic field, the $n_{\rm sk}=2$ SkX turns into another SkX, 
whose typical spin configuration is shown in Fig.~\ref{fig:skx}(b).
This is similar to the well-studied SkX with $n_{\rm sk}=1$, in which spins at the hull of each skyrmion are aligned parallel to the magnetic field. 
The $n_{\rm sk}=1$ state is well described by
\begin{eqnarray}
\!\!{\bf S}^{(n_{\rm sk}=1)}_i \!\!\propto\!\!\left(
    \begin{array}{c}
      \cos{\mathcal{Q}_{1 i}}-\frac{1}{2}\cos{\mathcal{Q}_{2 i}} -\frac{1}{2}\cos{\mathcal{Q}_{3 i}} \\
      \frac{\sqrt{3}}{2}\cos{\mathcal{Q}_{2 i}} -\frac{\sqrt{3}}{2}\cos{\mathcal{Q}_{3 i}}\\
      A_{1}(\sin \mathcal{Q}_{1 i}'  + \sin \mathcal{Q}_{2 i}' + \sin \mathcal{Q}_{3 i}') + \tilde{M}\!
    \end{array}
  \right)^{\!\!\!\!\rm T}\!\!,
  \label{eq:Qsk=1}
\end{eqnarray}
where $\mathcal{Q}_{\nu i}' = \mathcal{Q}_{\nu i} + \phi$ and ${\rm T}$ denotes the transpose of the vector;
 $A_1$, $\tilde{M}$, and $\phi$ are the variational parameters.
This is also a triple-$Q$ state: 
$S_{\bf q}$ has the peaks at ${\bf q}={\bf Q}_1$, ${\bf Q}_2$, and  ${\bf Q}_3$ with the same intensity, in addition to the induced ferromagnetic component at ${\bf q}=0$.
This is again viewed as a periodic array of a vortex with $v=-2$ and six merons with $v=1$, in a different manner from the $n_{\rm sk}=2$ SkX.

For a larger magnetic field, the spin state becomes topologically trivial, i.e., $n_{\rm sk}=0$.
The typical spin configuration  is shown in Fig.~\ref{fig:skx}(c), which is well approximated by another triple-$Q$ state: 
\begin{eqnarray}
{\bf S}^{(n_{\rm sk}=0)}_i \propto\left(
    \begin{array}{c}
      A_0\cos{\mathcal{Q}_{1 i}} + A^{\prime}_0\sin{\mathcal{Q}_{2 i}}\\
      A^{\prime}_0\sin{\mathcal{Q}_{1 i}'} - A_0\cos{\mathcal{Q}_{2 i}'}\\
      B_0\cos{\mathcal{Q}_{3 i}''} + \tilde{M}
    \end{array}
  \right)^{\rm T},
\label{eq:Qsk=0}
\end{eqnarray}
where $\mathcal{Q}_{\nu i}'' = \mathcal{Q}_{\nu i} + \phi'$; 
$A_0$, $A^\prime_0$, $B_0$, $\tilde{M}$, $\phi$, and $\phi'$ are the variational parameters. 
We note that the spin configuration is slightly different from 
those suggested in Refs.~\onlinecite{leonov2015multiply, PhysRevB.93.184413}.
As increasing the magnetic field, the triple-$Q$ components decrease and the $n_{\rm sk}=0$ state finally turns into a forced ferromagnetic state with $A_0 = A_0^\prime = B_0 = 0$ and $\tilde{M}=1$.

Figure~\ref{fig:skx}(d) summarizes the changes of $n_{\rm sk}$ 
and the magnetization in localized spins per site $M =|\sum_i {\bf S}_i|/N_s$ in an applied magnetic field $B$.
$n_{\rm sk}$ shows the two discontinuous changes: 
$n_{\rm sk} = 2 \to 1$ at $B_{\rm c1} \sim 0.00325$ and $n_{\rm sk}=1 \to 0$ at $B_{\rm c2} \sim 0.0065$.
These are successive topological transitions, similar to  
those characterized by the Chern number  
in the quantum Hall systems~\cite{arovas1984fractional}.
At the same time, $M$ changes discontinuously at the critical fields, while it 
grows almost linearly to $B$ in each phase.
Figure~\ref{fig:skx}(e) shows the $B$ dependence of the grand potential per site, $\Omega = E - \mu n$ ($E=\braket{\mathcal{H}}/N_s $ is the internal energy per site and $n=\sum_{i\sigma}\braket{\hat{c}^\dagger_{i\sigma}\hat{c}^{\;}_{i\sigma}}/N_s$ 
	 is the electron filling), obtained by the modified KPM-LD  simulation, in comparison with those for Eqs.~(\ref{eq:Qsk=2})-(\ref{eq:Qsk=0}) with optimal variational parameters~\cite{nsk2}. 
We see that the variational states give sufficiently precise energy in each $B$ range. 
We emphasize that the multiple digital changes of the topological number $n_{\rm sk}=2 \leftrightarrow 1 \leftrightarrow 0$ are very unique, as usually the $n_{\rm sk}=2$ SkX is hard to realize.

Let us discuss the energetics of the triple-$Q$ SkXs  
compared with the single-$Q$ states. 
As mentioned earlier, the RKKY interaction in the order of $J^2$, in general, prefers a helical state at zero field, which turns into a conical one for nonzero field, both of which are single-$Q$ states 
described by
${\bf S}_i 
\propto(\cos\mathcal{Q}_{\nu i}, \sin\mathcal{Q}_{\nu i}, \tilde{M} )$. 
The variational energy is plotted by the gray curve in Fig.~\ref{fig:skx}(e).
Our triple-$Q$ SkXs have a lower energy than the single-$Q$ state by $\sim 10^{-3}$. 
This energy difference is comparable to a higher-order term in the perturbation, $(J/2)^4S_{{\bf Q}_\nu}^2$, where $S_{{\bf Q}_\nu}\sim 1/6$ is the spin structure factor for these SkX states.
This confirms that the SkXs are stabilized by the higher-order contributions than the RKKY interactions. 
The details of the perturbative arguments 
will be reported elsewhere~\cite{preparation2}. 
 
In our simulation, we can directly obtain the electronic and magnetic properties of itinerant electrons.
It is worth noting that the electronic properties are obtained in a self-organized manner to optimize the grand potential of the system.
Figure~\ref{fig:electron} shows the real-space distributions of (a) the local solid angle $\Omega_p$ 
defined by $2{\bf S}_{p_1}\cdot({\bf S}_{p_2}\times{\bf S}_{p_3})/ \{ (\sum_{i=1}^3 {\bf S}_{p_i})^2-1 \}$, where ${\bf S}_{p_i}$ are three spins on a triangular plaquette $p$ in the counterclockwise direction,
(b) the local electric current density 
$j^\nu_{\langle i,j\rangle} = (1/2i)\sum_{\sigma}\braket{\hat{c}^\dagger_{i\sigma}\hat{c}^{\;}_{j\sigma} - \hat{c}^\dagger_{j\sigma}\hat{c}^{\;}_{i\sigma}}$ 
for a nearest-neighboring bond $\langle i,j\rangle$ 
in the $\nu$-direction [see the inset of Fig.~{\ref{fig:electron}}(b)],
(c) the length of itinerant electron spin 
$s_i = |\braket{\hat{\bf s}_i}|$, and (d) the local charge density
$n_i = \sum_{\sigma}\braket{\hat{c}^\dagger_{i\sigma}\hat{c}^{\;}_{i\sigma}}$.
Figures~\ref{fig:electron}(a)-\ref{fig:electron}(d)[\ref{fig:electron}(e)-\ref{fig:electron}(h)] show the modified KPM-LD results for the SkX  with $n_{\rm sk}=2$ ($1$)  at $B=0$ ($0.005$).  
$\Omega_p$ is positive for all the plaquettes and takes a $12$-sublattice superstructure for the $n_{\rm sk}=2$ SkX as shown in Fig.~{\ref{fig:electron}}(a), while it shows a  ferri-type $48$-sublattice superstructure for the $n_{\rm sk}=1$ SkX as shown in Fig.~{\ref{fig:electron}}(e). 
Reflecting the superstructures of $\Omega_p$, 
itinerant electrons form periodic lattices of circular currents surrounding the cores with small $|\Omega_p|$ plaquettes, as shown in Figs.~{\ref{fig:electron}}(b) and {\ref{fig:electron}}(f).
At the same time, electrons show the spin and charge density waves, as shown in Figs.~\ref{fig:electron}(c), \ref{fig:electron}(d), \ref{fig:electron}(g), and \ref{fig:electron}(h); 
$s_i$ and $n_i$ become larger in the region with smaller $|\Omega_p|$ and larger $j_{\langle i,j\rangle}^\nu$.
All the quantities have the threefold rotational symmetry, reflecting the spin textures 
in Figs.~\ref{fig:skx}(a) and~\ref{fig:skx}(b) (the symmetry is weakly lowered in Fig.~\ref{fig:electron}, presumably due to the statistical fluctuations)~\cite{mag}.

It will be interesting to observe the peculiar electronic properties by diffractions
or local probes in experiments. Furthermore, the spin and charge textures
of itinerant electrons are correlated with the localized spin textures as shown in Fig.~\ref{fig:electron}, 
suggesting the possibility of controlling the magnetism through the electronic channels.

Finally, we discuss another peculiar property of our SkXs. 
As the SkXs are stabilized by the Fermi surface effect through the peak structure of $\chi_{\bf q}^0$, we can control
the skyrmion size 
by the electron filling and band dispersion.
${\bf Q}_\nu$ are changed by $\mu$ and $t_{ij}$~\cite{suppl}.
This may lead to a change of the period of SkX in a wide range. 
Indeed, we confirm 
such a change by simulation. 
We note that, when the wave numbers  
are close to the zone boundary, 
the system may exhibit antiferromagnetic SkXs~\cite{zhang2016antiferromagnetic}.
Thus, our SkXs possess flexible controllability of the skyrmion size 
and the spin texture.

To summarize, we have theoretically discovered 
a SkX with unusually high topological number $n_{\rm sk}=2$ by large-scale unrestricted numerical simulation for the Kondo lattice model on a triangular lattice.
The SkX is stabilized at $T=0$ even in the absence of DM interactions and 
external magnetic field.
We also have found that the system exhibits unique successive topological transitions in applied magnetic field, with showing multiple digital changes of $n_{\rm sk} = 2 \to 1 \to 0$
and switching of itinerant electrons textures, such as circular electric currents and spin-charge density waves. 
Our discovery of the SkXs stabilized by the Fermi surface effect
may give an insight into exotic spin structures found in itinerant magnets, such as 
thin films on transition metal substrates~\cite{Heinze11}. 
Moreover, the flexibility in the multiple $n_{\rm sk}$,  skyrmion size, and helicity 
will be useful in potential applications.

\begin{acknowledgments}
We thank  C. D. Batista, Y. Kato, K. Barros, T. Misawa, and Y. Yamaji for fruitful discussions.
The KPM-LD simulations were carried out at the Supercomputer Center, 
Institute for Solid State Physics, University of Tokyo.
R.O. is supported by the Japan Society for the Promotion of Science through a research fellowship for young scientists and the Program for Leading Graduate Schools (ALPS). 
R.O. also acknowledges the CNLS summer student program at LANL.
 \end{acknowledgments}

\nocite{*}


\begin{thebibliography}{42}
\bibitem [{nsk()}]{nsk2}%
  \BibitemOpen
  \href@noop {} {}\bibinfo {note} {For the $n_{\rm sk}=2$ SkX in an applied field, we
  extend Eq.~(\ref{eq:Qsk=2}) by adding a net magnetization uniformly to all spin components for simplicity.}\BibitemShut {Stop}%
\bibitem [{mag()}]{mag}%
  \BibitemOpen
  \href@noop {} {}\bibinfo {note} {We note that the magnetic field breaks the threefold rotational symmetry of the $n_{\rm sk}=2$ SkX. The details will be discussed in Ref.~\onlinecite{preparation2}. }\BibitemShut {Stop}%
\end{thebibliography}
%

\end{document}